\documentclass[12pt]{iopart}

\usepackage{graphicx}
\def\spose#1{\hbox to 0pt{#1\hss}}

\def\lta{\mathrel{\spose{\lower 3pt\hbox{$\mathchar"218$}}
     \raise 2.0pt\hbox{$\mathchar"13C$}}}
\def\gta{\mathrel{\spose{\lower 3pt\hbox{$\mathchar"218$}}
     \raise 2.0pt\hbox{$\mathchar"13E$}}}
\newcommand{\be}{\begin{equation}}
\newcommand{\en}{\end{equation}}
\newcommand{\bea}{\begin{eqnarray}}
\newcommand{\ena}{\end{eqnarray}}

\begin{document}

\title[Static Vacuum Solutions in Non-Riemannian Gravity]{Static Vacuum Solutions in Non-Riemannian Gravity
}

\author{Rodrigo Maier}

\address{Institute of Cosmology and Gravitation, University of Portsmouth,\\
Dennis Sciama Building, Portsmouth, PO1 3FX, United Kingdom}
\address{Centro Brasileiro de Pesquisas F\'{\i}sicas -- CBPF, \\ Rua Dr. Xavier Sigaud, 150, Urca,
CEP 22290-180, Rio de Janeiro, Brazil}

\ead{rodmaier@cbpf.br}
\begin{abstract}
In the framework of non-riemannian geometry, we derive exact static vacuum
solutions of the field equations obtained from the full equivalent version of the Einstein-Hilbert action
when torsion degrees of freedom are taken into account.
By imposing spherical symmetry and a suitable choice for the contorsion degrees of freedom,
the static geometry provides deviations on the predictions of the observational tests predicted by General Relativity
-- namely on the advance of planetary perihelia and the bending of light rays --
which we infer. The analytical extension is built in two particular domains of the parameter space. In the first domain
we obtain a solution exhibiting an event horizon analogous to that of the Schwarzschild geometry.
For the second domain, we show that the metric furnishes an exterior event horizon,
and two interior horizons which enclose the singularity.
For both branches we examine the effects of torsion corrections on the Hawking radiation.
In this scenario the model extends Bekenstein's black hole geometrical thermodynamics, with an extra work
term connected to a torsion parameter.
\end{abstract}

\pacs{04.50.Kd; 04.20.Dw; 04.20.Jb}

%\pacs{98.80.Jk, 98.80.Qc, 05.45.-a}
\maketitle

\section{Introduction}
Although General Relativity is still most successful theory in order to describe gravitational interaction, there are crucial pathologies when one tries to obtain exterior static solutions engendered by gravitational collapse processes. Among such we can mention the prediction of singularities\cite{chandra}
-- sometimes plagued with instabilities on the Cauchy horizon\cite{starobinski} -- in black hole formation.

In General Relativity, black holes correspond to static solutions describing an exterior
spacetime of the final stage of gravitationally bounded systems whose masses
exceeded the limits for a finite equilibrium configuration\cite{chandra,wheeler}. Geometrically, a black
hole can be described as a region of asymptotically flat spacetimes bounded by
an event horizon hiding a singularity formed in the collapse. Fundamental theorems
by Israel and Carter\cite{israel,carter} state that the final stage of a general collapse of uncharged
matter is typically a Kerr black hole, which has an involved singularity structure.
Nevertheless, for a realistic gravitational collapse we have no evidence that the
Kerr solution describes accurately the interior geometry of the black hole.
One of the first theoretical solutions to this problem was furnished in the late 30's\cite{oppen},
indicating that the interior of the black hole, thus formed, is analogous to the interior of a Schwarzschild
black hole with a global spacelike singularity\cite{wald}. Although during the last
years many gravitational collapse models and numerical studies have been developed\cite{lake1}-\cite{luca}
the singularity issue still poses a problem in the final state of gravitational collapse processes.
As singularities cannot be empirically conceived, this turns out to be a pathology of Einstein's theory.

Despite the cosmic censorship hypothesis\cite{penrose},
there is no doubt that General Realtivity must be properly corrected or even
replaced by a completely modified theory of gravitation. In this direction,
loop quantum cosmology\cite{bojo} and string-based formalism of brane world
theory\cite{maier1,maier2} appear as attractive alternatives in order to solve
the singularity problem.

Even if a modified theory of gravitation would be able to circumvent the singularity issue,
the presence of Cauchy horizons in maximal analytical extensions (like in \cite{maier2}) still
poses an insurmountable question on the stability of spacetimes. In General Relativity, for instance,
the Cauchy horizon is a global boundary in which the field equations lose their power
to describe the evolution of prior initial conditions. It has been shown that for a free falling observer crossing the Cauchy horizon,
an arbitrary large blue-shift of any incoming radiation would be seen so that the
flux of energy of test fields would diverge once crossing it (see Ref. \cite{chandra} and references
therein). In this sense, the Cauchy horizon is a null surface of infinite blue-shift. As shown in \cite{starobinski}, this instability on the Cauchy horizon
is intrinsically related to free falling observers (and hence, related to the affine connection). Therefore, a modified theory of gravitation
which intends to avoid its divergences could, in principle, include torsion degrees of freedom.

Although there is still no current
observable evidences favouring torsion in gravity,
there are reasonable theoretical arguments for one to consider the introduction of
torsion fields as a desired component in a modified theory of gravitation\cite{niko}-\cite{hammond3}.
In cosmology, for instance, it has been shown\cite{niko} that the minimal coupling between Dirac spinors and
torsion generates a significant interaction at high energies, avoiding the undesirable initial singularity
which emerges from General Relativity. On the other hand, in string theory, the low-energy limit effective
Lagrangian has besides the aimed gravitational
field, a dilaton and an antisymmetric field \cite{hammond4,fradkin} in which
case the torsion potential can be an antisymmetric Kalb-
Ramond field. It has also been shown that if one wants to implement the
local Poincar\'e symmetry as part of a gauge theory then
torsion fields are also necessary (see \cite{shapiro}-\cite{hehl} for a review
on theories with torsion).

In order to perform a first analysis on the latter scenario, in this paper we seek for static vacuum solutions in the modified field equations derived from the full version of Einstein Hilbert action including torsion degrees of freedom.
In our search, we shall not be concerned with the
source of the torsion field. We will just consider the torsion as a
fundamental tensor, completely independent from the metric that defines the affine structure of spacetime.
In the next section we introduce the field equations together with our conventions.
In Section III, we restrict ourselves to a suitable choice of the torsion degrees of freedom
in order to obtain asymptotically flat exact solutions. The analytical extensions of our solutions
are exhibited in Section IV. In Section V we examine the corrections on the Hawking temperature
and employ our results in order the extend Bekenstein's
geometrical entropy. Conclusions and future perspectives are discussed in the final remarks.
\section{Field Equations}

Let ${g}_{\alpha\beta}$ be the metric tensor of a $4$-dimensional non-riemannian space endowed with a non-trivial affine structure due to torsion terms ${T}^{\alpha}_{~.~\beta\gamma}$. The connection can be defined as
\begin{eqnarray}
\label{eq1}
{\Gamma}^{\alpha}_{~\beta\gamma}= {\left\{^{\, \alpha}_{\beta\gamma}\right\}}+{K}^{\alpha}_{~.~\beta\gamma},
\end{eqnarray}
where ${\left\{^{\, \alpha}_{\beta\gamma}\right\}}$ are the Christoffel symbols and ${K}^{\alpha}_{~.~\beta\gamma}$ is the contorsion tensor. In this sense, the covariant derivative is defined as
\[
\nabla_{\beta}\xi^\alpha \equiv \xi^\alpha{}_{,\beta}+{\Gamma}^\alpha_{~\beta\gamma}\xi^\gamma.
\]
The torsion
\be
{T}^{\alpha}_{~.~\beta\gamma}\equiv {\Gamma}^{\alpha}_{~\beta\gamma}-{\Gamma}^{\alpha}_{~\gamma\beta}\label{eq2}
\en
together with the metricity condition  ${\nabla}_{\gamma}{g}_{\alpha\beta}=0$
allow us to write the contorsion as
\be \label{eq3}
{K}_{\alpha\beta\gamma}=\frac{1}{2}({T}_{\alpha\beta\gamma}+{T}_{\beta\alpha\gamma}+{T}_{\gamma\alpha\beta}),
\en
which has the anti-symmetry ${K}_{\alpha\beta\gamma}=-{K}_{\gamma\beta\alpha}$. For an arbitrary covariant vector field $Z_{\alpha}$, the curvature tensor ${R}^{\delta}_{~\alpha\beta\gamma}$ is then defined by
\be \label{eq3a}
\nabla_{\gamma}\nabla_{\beta}Z_{\alpha}-\nabla_{\beta}\nabla_{\gamma}Z_{\alpha}={R}^{\delta}_{~\alpha\beta\gamma}Z_{\delta}+T^{\delta}_{. ~\beta\gamma}\nabla_{\delta}Z_{\alpha},
\en
and can be separeted in its Riemannian and non-Riemannian parts as
\be\label{eq4}
{R}^{\alpha}_{~\beta\gamma\delta}={\tilde{R}}^{\alpha}_{~\beta\gamma\delta}+{K}^{\alpha}_{~\beta\gamma\delta}.
\en
Here, ${\tilde{R}}^{\alpha}_{~\beta\gamma\delta}$ is the Riemanian tensor defined only with the Christoffel symbols\cite{wald} and
\bea \label{eq5}
{K}^{\alpha}_{~\beta\gamma\delta}={D}_{\gamma}{K}^{\alpha}_{~.~\delta\beta}-{D}_{\delta}{K}^{\alpha}_{~.~\gamma\beta}+ {K}^{\mu}_{~.~\delta\beta}{K}^{\alpha}_{~.~\gamma\mu}
-{K}^{\mu}_{~.~\gamma\beta}{K}^{\alpha}_{~.~\delta\mu},
\ena
where ${D}$ is covariant derivative, constructed again, only with the Christoffel symbols.

In order to obtain the modified gravitational field equations, we assume the action
\bea\label{eq4r}
S=\frac{1}{2\kappa}\int\sqrt{-g}Rd^4x+\int {\cal L}_m(g_{\mu\nu},\Gamma^{\alpha}_{~\beta\gamma},\psi)d^4x,
\ena
where $g$ is the determinant of the geometry, $\kappa\equiv8\pi G$ and $R$ is the full Ricci scalar built with the metric plus contorsion degrees of freedom.
${\cal L}_m$ is the Lagrangian density of matter fields $\psi$ which, in general, also depends on torsion components.
We define the spin angular momentum tensor of matter as
\be
S_{\alpha}^{~~\beta\gamma}=-\frac{1}{\sqrt{-g}}\frac{\delta{\cal L}_m}{\delta K^{\alpha}_{~.~\beta\gamma}}\quad ,\label{spin}
\en

It can be shown\cite{clifton} that variations of $S$ with respect to $g^{\alpha\beta}$ and $T^\alpha_{~.~\beta\gamma}$, yield the complete set of field equations
\be
G_{\alpha\beta} \equiv \tilde{G}_{\alpha\beta}+{L}_{\alpha\beta}=C_{\alpha\beta}\quad ,\label{eq6}
\en
and
\be
T^{\alpha~~~\gamma}_{~.~\beta~.}+\delta^{\alpha}_{\beta}T^{\mu\gamma}_{~.~.~\mu}-g^{\alpha\gamma}T^{\mu}_{~.~\beta\mu}=\kappa S_{\beta}^{~~\alpha\gamma}\quad ,\label{eq6n}
\en
where $C_{\alpha\beta}$ is the canonical stress-energy tensor and
\bea
\tilde{G}_{\alpha\beta}\equiv {\tilde{R}}_{\alpha\beta}-\frac{1}{2}{\tilde{R}}\, {g}_{\alpha\beta} \quad , {L}_{\alpha\beta} \equiv {K}_{\alpha\beta}-\frac{1}{2}{K} \, {g}_{\alpha\beta} \label{eq8} \quad ,
\ena
with ${K}_{\alpha\beta}\equiv {K}^{\gamma}_{~\alpha\gamma\beta}$ and ${K}\equiv {g}^{\alpha\beta}~{K}_{\alpha\beta}$.

As in this analysis we shall not be concern with the source of the torsion field, we will assume that there is a lagrangian ${\cal L}_m$ so that equations (\ref{eq6n}) are automatically satisfied. Furthermore, in order to search for vacuum solutions, we will also impose that $C_{\alpha\beta}\equiv0$.

\section{Exact Solutions}

Let us assume a general static geometry with spherical symmetry in the coordinates $(t,r,\theta\phi)$ given by
\begin{eqnarray}
\label{eq3.1}
ds^2=F(r)dt^2-\frac{1}{G(r)}dr^2-r^2d\Omega^2,
\end{eqnarray}
where $d\Omega^2$ is the solid angle. By imposing the suitable {\it ansatz}
\begin{eqnarray}
\label{eq3.2}
K_{\alpha\beta\gamma}=(g_{\gamma\beta}\phi_{,\alpha}-g_{\alpha\beta}\phi_{,\gamma}),
\end{eqnarray}
together with the assumption $\phi\equiv\phi(r)$, we obtain an equivalent scalar-tensor theory
\begin{eqnarray}
\label{eq3.2n}
\tilde{G}_{\alpha\beta}=-L_{\alpha\beta}\equiv{T}_{\alpha\beta}
\end{eqnarray}
where the effective energy momentum tensor reads
\begin{eqnarray}
\label{eq3.2.1}
T_{\alpha\beta}=2\Big[g_{\alpha\beta}D_{\mu}D^{\mu}\phi-D_{\alpha}D_{\beta}\phi
-\Big(D_\alpha\phi D_\beta\phi+\frac{1}{2}g_{\alpha\beta}D_{\mu}\phi D^{\mu}\phi\Big)\Big].
\end{eqnarray}
It is straightforward to check that the general solution for (\ref{eq3.2n}) (or equivalently for (\ref{eq6}) with $C_{\alpha\beta}\equiv0$) is given by
\begin{eqnarray}
\label{eq3.4}
F(r)=e^{2\phi}\Big[1-\frac{2GM}{r}e^{\phi}\Big],~
G(r)=\frac{1}{(1-r\phi^\prime)^2}\Big[1-\frac{2GM}{r}e^{\phi}\Big].
\end{eqnarray}
Here we see that when $\phi\rightarrow 0$, the Schwarzschild solution is restored as one should expect.
The freedom in the scalar field comes from the fact that we have not fixed the source of the torsion field.
In this sense, the above vacuum solution could suggest an extension of the Birkhoff's theorem\cite{weinberg} when
the assumption (\ref{eq3.2}) is taken into account. It is worth mentioning that the conditions $D_{\alpha}L^{\alpha}_{~\beta}=0$
also hold so the Bianchi identities are automatically satisfied.

As in this first analysis we shall not concern about the source of torsion degrees of freedom, we now proceed
by investigating asymptotically flat solutions in order to fix the scalar function $\phi$.
Although there might be several functions which satisfy this requirement, from now on we are going to restrict ourselves
to a particular choice which provides a simply deviation from the Schwarzschild geometry through a special
parameter $\alpha$. That is, by fixing
\begin{eqnarray}
\label{eq3.7}
\phi(r)=\ln\Big|1+\frac{\alpha}{r}\Big|,
\end{eqnarray}
the geometry (\ref{eq3.1}) reads
\begin{eqnarray}
\label{eq3.8}
\nonumber
ds^2=\frac{(\alpha+r)^2[r^2-2GM(r+\alpha)]}{r^4}dt^2\\
~~~~~~~~~~~~~~-\frac{r^2(2\alpha+r)^2}{(\alpha+r)^2[r^2-2GM(r+\alpha)]}dr^2-r^2d\Omega^2.
\end{eqnarray}
which is asymptotically flat. It is worth mentioning that
although we should not focus on the source content in this first analysis, here we assume that there exists a Lagrangian ${\cal L}_m$ which can meet the simple choice (\ref{eq3.7}).

When $\alpha\rightarrow 0$ ($\phi\rightarrow 0$) we recover the Schwarzschild
solution as one should expect. In this sense, the small constant parameter $\alpha$ would make explicit any deviations from
the Schwarzschild geometry, and hence, from the observational tests of General Relativity.
In fact, expanding (\ref{eq3.8}) up to first order in $\alpha$ and holding
only the terms up to first order in $r^{-1}$, one may show that the advance of planetary perihelia per revolution
is given by
\begin{eqnarray}
\label{eqpa}
\Omega=\frac{6\pi(GM+\alpha)}{a(1+e^2)}
\end{eqnarray}
where $e$ is the eccentricity of the orbit and $a$ its semi-major axis. Analogously, for the bending of light rays passing in the neighborhood of a spherical
massive body, the deflection angle of the asymptotes is given by
\begin{eqnarray}
\label{eqpb}
\Delta \varphi=\frac{4(GM+\alpha)}{R}
\end{eqnarray}
where $R$ is the radius of the body. In this sense, both predictions could be, in principle, tested for a sufficiently small value of
$\alpha$ compared to $GM$. For $\alpha=0$, we obviously recover the predictions of General Relativity.

Depending on the domain of $\alpha$, two branches of analytical extensions emerge.
In fact, in the next section we will see how the solution (\ref{eq3.8}) bifurcates
in each branch so that the sign of $\alpha$ drastically changes the structure
of the spacetime.

\section{Maximal Analytical Extensions}

In order to examine the analytic completion of the exterior geometry (\ref{eq3.8}), we need
to know whether, and under what circumstances, the configuration forms event
horizons. By defining the polynomial
\begin{eqnarray}
\label{eqi2n}
P(r)=(\alpha+r)^2[r^2-2GM(r+\alpha)]
\end{eqnarray}
we see through an immediate inspection that (\ref{eq3.8}) does not allow naked singularities
configurations. In fact, it is straightforward to show that the roots of $P(r)$ are given by
\begin{eqnarray}
\label{eqeh1}
R_{in}&=&-\alpha\\
R_-&=&GM-\sqrt{GM(GM+2\alpha)}\\
R_+&=&GM+\sqrt{GM(GM+2\alpha)}.
\end{eqnarray}

If $\alpha>0$, the geometry (\ref{eq3.8}) provides one event horizon
-- analogous to that of the Schwarzschild solution -- given by $R_+$.
On the other hand, for $-GM/2<\alpha<0$ we obtain a similar exterior
event horizon $R_+$ together with two interior horizons $R_{in}$
and $R_-$, with $R_{in}<R_-$. As from the physical point of view $\alpha$
should be a small parameter, in the following we are going to restrict
ourselves to $|\alpha|<GM/2$. In this case, the deviation from General Relativity
in the Schwarzschild event horizon is given by
\begin{eqnarray}
\label{eqeh1n}
R_+=2GM+\alpha.
\end{eqnarray}
up to first order in $\alpha$, in both domains $+GM/2>\alpha>0$ and $-GM/2<\alpha<0$.

To proceed with the analytical extension of (\ref{eq3.8}), let us consider the following coordinates transformation
\begin{eqnarray}
\label{eqi1}
\frac{2\gamma}{u}du:=-\frac{r^3 (2\alpha + r)}{P(r)}dr-dt,\\
\frac{2\gamma}{v}dv:=-\frac{r^3 (2\alpha + r)}{P(r)}dr+dt.
\end{eqnarray}
Therefore, the geometry (\ref{eq3.8}) can be rewritten as
\begin{eqnarray}
\label{eqi3}
\nonumber
ds^2=-\frac{4\gamma^2}{uv}\frac{P(r)}{r^4}dudv-r^2d\Omega^2.
\end{eqnarray}
By defining
\begin{eqnarray}
\label{eqi4}
r^{\ast}:=\int \frac{r^3 (2\alpha + r)}{P(r)} dr,
\end{eqnarray}
we obtain
\begin{eqnarray}
\label{eqi5n}
r^{\ast}=r + \frac{\alpha^2}{\alpha + r} - 2 GM \ln|\alpha + r|+ 2 GM \ln|r^2-2GM( \alpha + r)|
\end{eqnarray}
so that integration of (\ref{eqi1}) yields
\begin{eqnarray}
\label{eqi7}
r^{\ast}=-\gamma\ln|uv|~~,~~t=\gamma\ln|v/u|.
\end{eqnarray}
\begin{figure}
\begin{center}
\includegraphics[width=8cm,height=6cm]{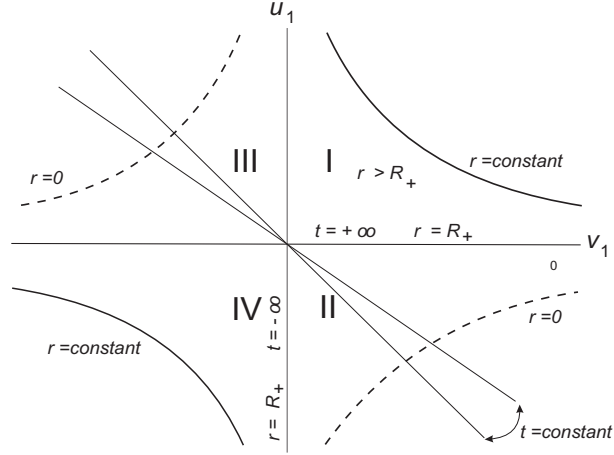}
\caption{The Kruskal extension of the spacetime (\ref{eq3.8}) for $\alpha>0$.}
\end{center}
\end{figure}
\begin{figure}
\begin{center}
\includegraphics[width=8cm,height=6cm]{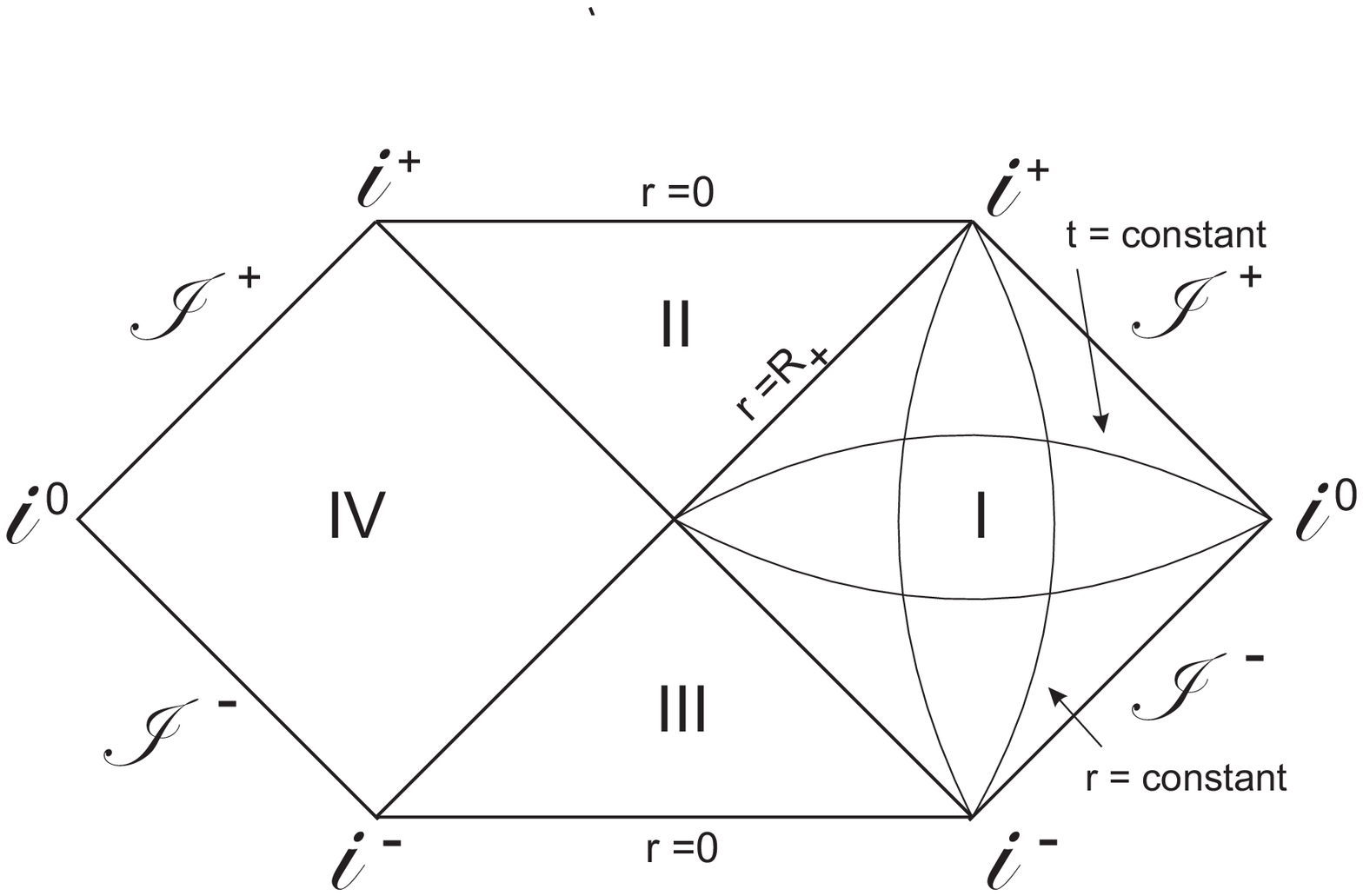}
\caption{Penrose diagram for the analytical extension of the spacetime (\ref{eq3.8}) for $\alpha>0$. Here, the event horizon is given by $R_+=GM + \sqrt{GM( GM + 2 \alpha  )}$}
\end{center}
\end{figure}

For the case $\alpha>0$, one may consider the chart $(u_1,v_1)$ by setting $\gamma=-2GM$. From (\ref{eqi7}) we find that
\begin{eqnarray}
\label{eqi9n}
u_1v_1=\Big[\frac{r^2-2GM(\alpha+r)}{\alpha+r}\Big]e^{\frac{1}{2GM}\Big[r+\frac{\alpha^2}{\alpha+r}\Big]},
\end{eqnarray}
and this provides a regular covering for any subregion with $r>0$. Figure 1 is the Kruskal-type diagram which give a faithful map of any subregion covered by the chart $(u_1,v_1)$. In this case, the maximal analytical extension of spacetime (\ref{eq3.8}) is analogous
to that of a Schwarzschild black hole\cite{chandra} with an event horizon
$R_{+}$ (cf. Fig. 2).

If $-\frac{GM}{2}<\alpha<0$, one may
consider again the chart (\ref{eqi9n}) for $r>R_{in}$. In this case we see
that the metric exhibits no singularity at
$R_{-}$ and $R_{+}$. However, the chart (\ref{eqi9n}) does not furnish a regular covering for any subregion $r<R_{-}$.
In order to circumvent this problem, one may go to the chart $(u_2,v_2)$
by setting $\gamma=2GM$. In this case we obtain
\begin{eqnarray}
\label{eqi8}
\nonumber
u_{2}v_{2}=\Big[\frac{\alpha+r}{r^2-2GM(\alpha+r)}\Big]e^{-\frac{1}{2GM}\Big[r+\frac{\alpha^2}{\alpha+r}\Big]},
\end{eqnarray}
so that (\ref{eq3.8}) exhibits no singularity at $r=R_{in}$. In fact, the chart $(u_{2},v_{2})$ gives a regular mapping
of any given subregion of the manifold which has $r<R_{-}$.
In the domain of overlap $R_{in}<r<R_{-}$ the two charts are related by
\begin{eqnarray}
\label{eqi10}
|u|_{1}=|u|^{-1}_2~~,~~|v|_{1}=|v|^{-1}_2.
\end{eqnarray}
In Fig. 3 we show a fundamental portion of the maximal analytical extension of spacetime (\ref{eq3.8}) for $-\frac{GM}{2}<\alpha<0$. Here we see that it exhibits an exterior event horizon $R_{+}$,
together with two interior horizons $R_{-}$ and $R_{in}$ ($R_{-}>R_{in}$) enclosing the singularity at $r=0$.

\begin{figure}
\begin{center}
\includegraphics[width=8cm,height=9.5cm]{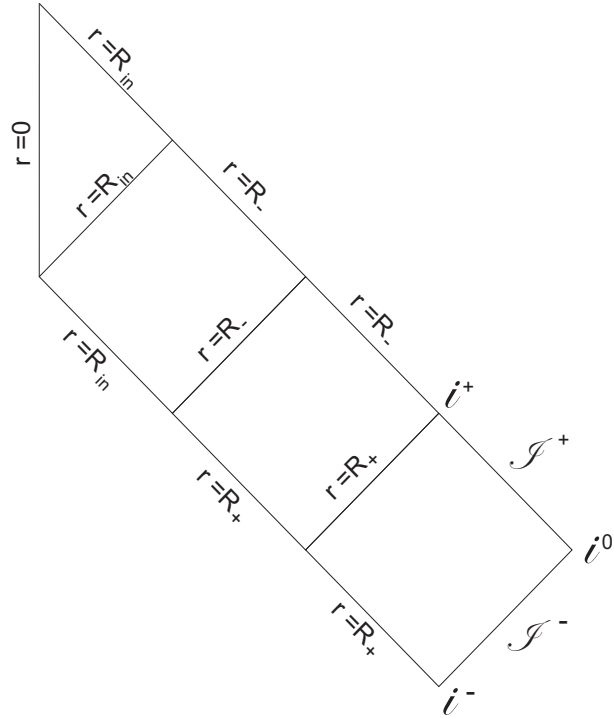}
\caption{Penrose diagram for the case $-\frac{GM}{2}<\alpha<0$. Here we see that the spacetime exhibits an exterior event horizon $R_{+}$ and two interior horizons
$R_{-}$ and $R_{in}$. The complete analytical extension can be obtained by connecting asymptotically
flat regions, like the fundamental portion shown above, in an infinite chain.}
\end{center}
\end{figure}

From the mathematical point of view, it is worth mentioning though that in the case of $\alpha=-\frac{GM}{2}$, we obtain a solution analogous to that of the Reissner-Nordstrom geometry. It can also be shown that for $\alpha<-GM/2$, the analytical extension is analogous to that of the Schwarzschild metric.

\section{Hawking Radiation and Black Hole Thermodynamics}

Using a semi-classical approach, S. W. Hawking derived the thermal spectrum of emitted particles by a black hole\cite{hawking}.
Following his same original procedure, we now intend to obtain the correction on the Hawking radiation when the torsion
degrees of freedom are introduced through (\ref{eq3.8}).
To do so, let us then consider a test massless Klein-Gordon field $\zeta$ in the background defined by the spacetime (\ref{eq3.8}).
The propagation of scalar field is then described by the scalar wave equation
\begin{eqnarray}
\label{eqVI2}
g^{\alpha\beta}D_{\alpha}D_\beta \zeta=0.
\end{eqnarray}
Given the symmetries of the background we search for a solution as
\begin{eqnarray}
\label{eqVI3}
\zeta_{\omega ml}=\frac{1}{r} R_{\omega l}(R^{\ast}) Y_{ml}(\theta, \phi) \exp (-i\omega t).
\end{eqnarray}
where
\begin{eqnarray}
\label{eqVI3nn}
\nonumber
R^{\ast}=\int\frac{1}{G(r)}dr= \frac{r^2+\alpha r-\alpha^2}{\alpha+r}+2\{(\alpha+GM)\ln|r+\alpha|\\
\nonumber
+\sqrt{GM (2 \alpha + GM)} [\ln|r-GM- \sqrt{GM (2 \alpha + GM)}|  \\
- \ln|r-GM + \sqrt{GM (2 \alpha + GM)}|]\}.~~~~~~~~~~
\end{eqnarray}
Using (\ref{eqVI3}) and (\ref{eqVI3nn}), the wave equation (\ref{eqVI2}) is reduced to an ordinary differential equation
in $R^{\ast}$. It is straightforward to see that in the asymptotical limit $r\rightarrow\infty$, this equation reduces to
\begin{eqnarray}
\label{eqVI5}
\nonumber
\frac{d^2R_{\omega l}}{d{R^{\ast}}^2}+ \omega^2R_{\omega l}=0 \Rightarrow R_{\omega l}(R^{\ast})=\exp(\pm i\omega{R^{\ast}}).
\end{eqnarray}
Therefore one may write the asymptotical Klein Gordon field as
\begin{eqnarray}
\label{eqVI6}
\zeta_{~1}=\frac{1}{r} \exp[-i\omega(t-{R^{\ast}})] Y_{ml}(\theta, \phi)
\end{eqnarray}
and
\begin{eqnarray}
\label{eqVI7}
\zeta_{~2}=\frac{1}{r} \exp[-i\omega(t+{R^{\ast}})] Y_{ml}(\theta, \phi).
\end{eqnarray}
\par
Let us now assume that the source that generates the exterior solution (\ref{eq3.8}) is given by a thin shell of a spherically symmetric matter distribution, and the flat spacetime inside such distribution is given by
\begin{eqnarray}
\label{eqVI8}
ds^2=dT^2-dr^2-r^2d\Omega^2.
\end{eqnarray}
Defining $r=R(t)$ as the scale factor that describes the evolution of the matter distribution, we impose that the interior metric match the exterior geometry by the following equation
\begin{eqnarray}
\label{eqVI9}
1-\Big(\frac{dR}{dT}\Big)^2=F(r)\Big(\frac{dt}{dT}\Big)^2-\frac{1}{G(r)}\Big(\frac{dR}{dT}\Big)^2.
\end{eqnarray}
We also define the respective null interior and exterior coordinates by
\begin{eqnarray}
\label{eqVI10}
V:=T+r~,~~U:=T-r~,
\end{eqnarray}
and
\begin{eqnarray}
\label{eqVI11}
v:=t+R^{\ast}~,~~u:=t-R^{\ast}.
\end{eqnarray}

Assuming that the null incident rays reach the matter distribution when $r=R_I\gg R_{+}$ (that is, $F(r)\sim1$, $G(r)\sim1$), we obtain
\begin{eqnarray}
\label{eq52}
\Big(\frac{dT}{dt}\Big)^2\simeq 1\rightarrow t\simeq T.
\end{eqnarray}
Thus,
\begin{eqnarray}
\label{eq53}
v_I\simeq T+R^{\ast}\Rightarrow V_I=v_I-\kappa
\end{eqnarray}
where
\begin{eqnarray}
\label{eq54}
\kappa=R_I-R^{\ast}(R_I).
\end{eqnarray}
\par
When $r=0$, we derive the trivial relation between $V$ and $U$ at the center of the matter distribution:
\begin{eqnarray}
\label{eq55}
V_0=T=U_0.
\end{eqnarray}
\par
Let us now consider that the outgoing waves emerge from the matter distribution when $r=R_{II}\sim R_{+}$. If
$T_{0}$ is taken to be the instant in which $r=R_{+}$, one may expand the scale factor $R_{II}(T)$ in Taylor series as
\begin{eqnarray}
\label{eq56}
R_{II}(T)\simeq R_{+} + A(T_{0}-T),
\end{eqnarray}
where $A$ is a constant.
Therefore, from equation (\ref{eqVI9}) we have
\begin{eqnarray}
\label{eq57}
t\simeq - 2GM\ln|T_0-T|.
\end{eqnarray}
However, from (\ref{eqVI3nn})
\begin{eqnarray}
\label{eq58}
\nonumber
R^{\ast}\simeq 2\sqrt{GM (2 \alpha + GM)} ~~~~~~~~~~~~~~~~~~~~~~~~\\
\times\ln|r-GM- \sqrt{GM (2 \alpha + GM)}|.
\end{eqnarray}
Then we obtain
\begin{eqnarray}
\label{eq59}
u_{II}&\simeq&-\delta\ln|T_{0}-T|,\\
U_{II}&\simeq &\chi \exp{\Big(-\frac{u_{II}}{\delta}\Big)}+\psi,
\end{eqnarray}
where $\delta\equiv2R_+$, and
\begin{eqnarray}
\label{eq61}
\chi=-(1+A)~,~~\psi=T_{0}-R_{+}.
\end{eqnarray}
As $U_0=V_0$ at $r=0$, the relations between the exterior null coordinates are given by
\begin{eqnarray}
\label{eq63}
v=v_{0}+\chi\exp{\Big(-\frac{u}{\delta}\Big)}~,~~u=-\delta\ln{\Big|\frac{v-v_{0}}{\chi}\Big|},
\end{eqnarray}
with $v_{0}\equiv\psi+\kappa$.
\par
Using (\ref{eqVI7}) we now expand $\zeta_{1\omega lm}$ in terms of $\zeta_{2\omega lm}$ as
\begin{eqnarray}
\label{eq641}
\nonumber
\varphi_{1\omega lm}= \int^\infty_0 [\alpha^{\ast}_{\omega^{\prime} \omega l m} \exp{(-i\omega^{\prime} v)}
-\beta_{\omega^{\prime} \omega l m} \exp{(i\omega^{\prime} v)}]d\omega^{\prime}.
\end{eqnarray}
Here, $\alpha^{\ast}_{\omega^{\prime} \omega l m}$ and $\beta_{\omega^{\prime} \omega l m}$ are the so-called Bogolubov coefficients\cite{birrel} so that
it is straightforward to show\cite{hawking} the relation
\begin{eqnarray}
\label{eq644}
|\alpha_{\omega^{\prime} \omega l m}|=e^{\pi \omega \delta} |\beta_{\omega^{\prime} \omega l m}|.
\end{eqnarray}
Furthermore, it follows from the orthogonality property of $\zeta_{1\omega lm}$ and $\zeta_{2\omega lm}$ that
\begin{eqnarray}
\label{eq645}
\sum_{\omega^{\prime}} [~|\alpha_{\omega^{\prime} \omega l m}|^2 - |\beta_{\omega^{\prime} \omega l m}|^2~]=1.
\end{eqnarray}
Using (\ref{eq644}) and (\ref{eq645}) we then obtain that the spectrum of the average number of created particles on the $\omega lm$ mode is given by
\begin{eqnarray}
\label{eq41rr}
N_{\omega lm}=\sum_{\omega^{\prime}} |\beta_{\omega^{\prime} \omega l m}|^2 =\frac{1}{\exp{(2\pi\delta\omega)}-1}.
\end{eqnarray}
The above result corresponds to a Planckian spectrum with associated temperature
\begin{eqnarray}
\label{eq42rr}
T_H=\frac{1}{2 \pi \delta}.
\end{eqnarray}
\par
Here we see that the Hawking temperature depends on the parameter $\alpha$
connected to torsion degrees of freedom.
In this sense, the observation of Hawking radiation could, in principle,
allows us to test our results for a finite and small value of $\alpha$.

Another feature which demands a carefully analysis is related to the entropy.
Motivated by the analysis of energy processes involving black holes, Bekenstein\cite{beken}
made the remarkable assumption that the entropy of a black hole should be proportional
to the area of its event horizon and formulated a first law of black hole thermodynamics.
According to his first law, the surface gravity of the black hole appear as proportional to a temperature.
Bekenstein's results, however, are rather geometrical and
did not involve any fundamental principle of statistical mechanics. Two years later,
by examining the quantum creation of particles near a Schwarzschild
black hole, Hawking showed that the black hole emits particles with a Planckian thermal
spectrum of temperature $T \propto \kappa$ where $\kappa$ is the surface gravity of the black hole. This striking result fits exactly
in the Bekenstein formula for the first law of black hole thermodynamics\cite{bardeen}, thus validating
Bekenstein's geometrical proposals and fixing the proportionality factor connecting the
entropy and the area of the black hole.

In the case of our solution (\ref{eq3.8}), it can be shown that the results of geometrical black hole thermodynamics
of Bekenstein might be extended. In fact, let us consider the expansion of $R_+$ up to first order in $\alpha$
according to (\ref{eqeh1n}).
Denoting $A_{eh}$ as the outer horizon spherical area, we obtain
\begin{eqnarray}
\label{eqb2}
dA_{eh}\simeq8\pi R_{+}(2GdM+d\alpha),
\end{eqnarray}
or
\begin{eqnarray}
\label{eqb3}
\frac{1}{4G}dA_{eh}\simeq\frac{1}{T_H}\Big(dM+\frac{1}{2G}d\alpha\Big)
\end{eqnarray}
according to (\ref{eq42rr}). We can therefore associate the outer horizon area of the
black hole with the geometrical entropy
\begin{eqnarray}
\label{eqb4}
S_{geom}=\frac{1}{4G}A_{eh},
\end{eqnarray}
a result which, in units $\hbar=1$, $K_B=1$, is in accordance to Bekenstein's definition\cite{beken}. Equation (\ref{eqb3}) is an
extended First Law with an extra work term connected to torsion degrees of freedom through the parameter $\alpha$. For deviations with $\alpha=const.$, we recover the form of the first law for the Schwarzschild black hole

\section{Final Remarks}

In this paper we examined static vacuum solutions of the full Einstein-Hilbert action when torsion
degrees of freedom are taken into account. Choosing a suitable form for the contorsion components,
we obtain arbitrary general solutions so that we fix the scalar field in order to provide
asymptotically flat configurations. The solutions obtained
do not allow for naked singularities in any domain of the parameter space. In this sense, the cosmic censorship hypothesis\cite{penrose} still holds
in our scenario.
In fact, the structure of spacetime bifurcates in two main
particular branches. In the first domain, the analytical extension is analogous to that of the Schwarzschild
solution with an event horizon. For the second branch, we show that the geometry provides an exterior event horizon,
and two interior horizons which enclose the singularity. Comparing to the Reissner-Nordstrom geometry, the interior horizon $R_-$ resembles a Cauchy
horizon in the sense that $P(r)<0$ for $R_-<r<R_+$, and $P(r)\geq0$ for $r<R_-$ (cf. Eq. (\ref{eqi2n})). If that is the case,
the presence of $R_-$ would pose a question on the stability of the spacetime (\ref{eq3.8}) for $-GM/2<\alpha<0$. Due
to its similarity to the Reissner-Nordstrom space–time, we should expect that the
flux of energy of test fields may diverge on crossing $R_-$. However, a complete treatment of the problem
should include the torsion degrees of freedom together with higher order nonlinear terms so as to provide sufficient conditions for instability. We will
address these issues in a future publication.

The aim of the analysis performed in this paper was to seek for static vacuum solutions
in non-riemanian gravity in order to better understand how torsion degrees of freedom may provide deviations from
General Relativity. In this sense, our central result shows how the detection of torsion
might be related to small deviations of the Schwarzschild horizon.
Furthermore, we also examine the effects of torsion corrections on the
observational tests of General Relativity and on
Hawking radiation as simple applications. In the case of the advance of planetary perihelia
and the bending of light rays, our predictions could be tested at least in low energy regimes for a sufficiently small value of
$\alpha$ compared to $GM$. On the other hand, the
calculation of the modified Hawking temperature allowed us to derive, analogously
to Bekenstein, a geometric entropy that confirms
the classical prediction that the entropy is proportional to the area of the event
horizon. Although the classical black hole thermodynamics
introduced by Bekenstein was validated by Hawking's semiclassical
derivation of the black body thermal emission of a black hole, black
hole thermodynamics always seemed to possess a heuristic character since no basic
principle of statistical mechanics was used in its derivation. Indeed, the definition
of the entropy of black holes is still an open problem and we actually refer to it as
a geometrical entropy.

In the framework of our general solutions, some remaining issues still need a more careful analysis.
The first is to determine what would be the internal solution given by a fluid matter distribution
which could be matched, and hence engender, the external vacuum solutions (\ref{eq3.8}). The lagrangian
${\cal L}_m$ to be introduced in (\ref{eq4r}) to meet (\ref{eq3.7}),
is another feature which deserves a more careful attention in order to better understand
the source of torsion fields for (\ref{eq3.8}). These issues together with a more comprehensive analysis of the
interior horizons $R_-$ and $R_{in}$ will be subject of a future paper.

\section{Acknowledgements}
RM acknowledges financial support
from CNPq/MCTI-Brasil, through a post-doctoral research grant no. 201907/2011-9 and BEV research grant no. 170047/2014-8.
RM also acknowledges the Institute of Cosmology and Gravitation, University of Portsmouth, and the Centro Brasileiro de Pesquisas
F\'isicas for their hospitality.

\section*{References}

\end{document}